\documentstyle[11pt]{article}
\newcommand{\be}{\begin{equation}}
\newcommand{\ee}{\end{equation}}
\begin{document}
\large
\title{\bf The \v Cerenkov-Compton effect in particle physics}
\author{\bf Miroslav Pardy\\
{Department of Theoretical Physics and Astrophysics}\\
{Masaryk University, Kotl\'{a}\v{r}sk\'{a} 2, 611 37 Brno}\\
{Czech Republic}\\
{E-mail: pamir@physics.muni.cz}}
\date{\today}
\maketitle
\baselineskip 15pt
\begin{abstract}
The two-photon production by motion of a charged particle in a medium is considered.
This process is called \v Cerenkov-Compton effect  (\v CCE) because the
production of photons is calculated from the Feynman diagram for the Compton effect in a medium.
This process is not forbidden in quantum electrodynamics of dielectric
media.The cross section of this process in Mandelstam variables
is calculated for pair of photons with one moving at the  opposite direction to the electron motion
and the second one inside \v Cerenkov cone. The opposite motion is not caused by the collision with a particle of
a medium. The relation of this process to the CERN experiments is considered.
\end{abstract}
\hspace{3ex}
\large

\baselineskip 14pt
\section{Introduction}

The \v Cerenkov-Compton effect (\v CCE) is in this article considered as
the synergic two-photon production by a charged particle moving in
a medium in such a way that it has some analogy of the \v Cerenkov
and Compton mechanism.
The two-photon process caused by the \v Cerenkov mechanism was considered by Frank
in 1968 as the anomalous scattering of light on a particle in medium [1].
In the Frank formulation of the  process, the initial photon
induces the final photon. The total process is then the two-photon
emission  in a
medium. The classical theory of this process was elaborated by Frank and
Tsytovich in 1980 [2]. The theory of these authors is not based on the quantum field theory and Feynman diagrams.

The \v Cerenkov mechanism requires the velocity of particle is greater than the velocity of light in the medium.
It can be determined classically supposing some microscopical structure of medium
and interaction of moving charges with this microscopical structure [3].
On the other hand, it can be determined quantum mechanically
using Feynman diagram where it is not used the specific
microscopical mechanism of medium for calculation of this effect [4].
Only the index of refraction is postulated. Also Schwinger source
theory works without microscopical mechanism generating radiation.
Our problem differs from the Frank problem in a sense,
that the emission of two photons is here considered as the synergic process
of the \v Cerenkov mechanism  and the Compton one.

We will show that this process is not forbidden in the framework of quantum field theory.
We describe \v CCE in terms of quantum electrodynamics.

We use the Feynman diagram for the Compton process as an extension and analogue of the
Feynman diagram for the simple \v Cerenkov process, in order to describe it.
So,  we do not use the knowledge of the microscopical mechanism of this process.

This process was not discussed in textbooks and in
monographies [3] on the \v Cerenkov radiation. On the other hand, it
seems that there is an experimental evidence of this effect in CERN,
because every photography with the \v Cerenkovian circles involves
the background with photons which are probably generated by the \v CCE mechanism.

While \v Cerenkov process has the symbolic description

$$e \longrightarrow e + \gamma\eqno(1),$$
the Compton process has the symbolic description

$$e  + \gamma  \longrightarrow e + \gamma\eqno(2).$$

The \v Cerenkov-Compton process follows from the last symbolic equation as
the transposition of its components and has the symbolic expression in the form:

$$e \longrightarrow e + \gamma + \gamma.\eqno(3)$$

This process is physically allowed for the index of refraction $n > 1$ and for
a charged particle moving through the transparent medium at
speed greater than the speed of light in that medium.

The rigorous
description of the physical reality of this process cannot be realized  by some mechanical model,  but
only by means of Feynman  diagrams with the corresponding
mathematical theory of the Compton process. So, in other words, while the ordinary \v Cerenov effect enables
mechanical explanation, as it was shown by Tamm and Frank [6], the origin of the \v CCE is quantum electrodynamical.

Let us recall some ingredients concerning \v Cerenkov radiation.
The so called \v Cerenkov radiation was observed
experimentally first by \v{C}erenkov [5] and theoretically
explained by Tamm and Frank [6] in classical electrodynamics as
a shock wave resulting from a charged particle moving through a transparent
material faster than the velocity of light in the material.
The source theory explanation was given by
Schwinger et al. [7] and the particle production by the \v{C}erenkov
mechanism was discussed by Pardy [8,9]. The \v Cerenkov effect at finite
temperature in source theory was discussed in [10,11] and
the \v Cerenkov effect with radiative corrections, in electromagnetism and gravity was analysed in [12,13].

The \v Cerenkov effect in source theory follows from amplitude

$$\langle 0_{+}|0_{-} \rangle  = e^{\frac {i}{\hbar} W},\eqno(4)$$
where $W$ is the action of electromagnetic field in medium. It seams that
Feynamn diagram methods are more appropriate to describe this process.

In order to be pedagogically clear, we derive in section 2
some elementary energetical and angle relations concerning
the \v CCE. Then, in the following section 3 we
prove the theorem on the existence of the \v CCE. In section 4,
the cross-section of the \v CCE is derived.  Discussion in section
5 is devoted to the summary and the experimental evidence of \v CCE in CERN.

\section{Energetical and angle relations}

In the pure \v Cerenkov process the four-momentum is of the form

$$p = k + p'.\eqno(5)$$

In the pure Compton process,
the four-momentum relation of the initial and final parameter is as follows:

$$p + k = p' + k.'\eqno(6)$$

However, in case of the two-photon \v Cerenkov process which we denote
as the \v Cerenkov-Compton effect, the four-momentum relation is of the form:

$$p = k + k' + p'.\eqno(7)$$

This equation can be splitted into two equations for energy and momentum:

$$E = \omega + \omega' + E'\eqno(8)$$
and

$${\bf p} = {\bf k} + {\bf k}' + {\bf p}'.\eqno(9)$$
where $\omega$ and $\omega'$ are energies of emitted photons and $E$ and
$E'$ are initial and final energies of an electron, ${\bf k}$ and ${\bf k}'$
are wave vectors of the emitted photons and ${\bf p}$ and ${\bf p}'$
are initial and final three momenta of electron. We suppose in this
article that $c = \hbar = 1$.

Now, let us introduce angle between ${\bf k}$ and ${\bf k'}$ as $\varphi$,
an angle between {\bf p} and {\bf k} as $\chi$ and angle between ${\bf p'}$
and ${\bf k'}$ as $\alpha$, or, $\angle ({\bf k},{\bf k'}) = \varphi,
\angle({\bf p},{\bf k}) = \chi, \angle({\bf p'},{\bf k'}) = \alpha$.

We can derive simple energy equation using the procedure of Di  Giacomo et al. [4]:

$$ \omega + \omega' =  E({\bf p}) - E({\bf p}')  =
E({\bf p}) - E({\bf p} - ({\bf k} + {\bf k}')) =
\frac {\partial E}{\partial \bf p}
({\bf k} + {\bf k}') = {\bf v}({\bf k} + {\bf k}'), \eqno(10)$$
or, ($k = \omega n, k' = \omega' n$)

$$\omega\cos\chi + \omega'\cos\alpha = \frac{\omega + \omega'}{vn}.
\eqno(11)$$

From eq. (11) it follows that for $\alpha = \pm \chi$ we get the \v Cerenkov threshold
$\cos \chi = 1/nv$. It means that the total \v Cerenkovian  spectrum
involves the ordinary \v Cerenkovian spectrum of photons and some part of the photon spectrum of the \v CCE.

Eliminating $\cos \varphi$ from the last equation we get:

$$\cos\varphi = \frac {\cos\alpha}{vn} + \frac {\omega'}{\omega}
\left(\frac {\cos\alpha}{vn} - \cos^{2}\alpha\right) +
\sin\alpha\sqrt{1 - \left(\frac {1}{vn} + \frac {\omega'}{\omega}
\left(\frac {1}{vn}-\cos\alpha\right)\right)^{2}}.\eqno(12)$$

For the sake of simplicity, let us consider the experimental situation
with the backward emission of photon with frequency $\omega'$, or,with $\alpha = \pi$.

Using  $ \varphi = \chi-\pi$,  and
eq. (11), or (12) we get the following equation:

$$\cos\chi = \frac {1}{vn} +
\frac {\omega'}{\omega}\left(\frac {1}{vn} + 1\right).\eqno(13)$$

We see, that the last equation involves the \v Cerenkov threshold
$\cos\chi = 1/nv$ and it means that the relation of the considered effect to the \v Cerenkov effect is appropriate.

From eq. (13) the following relations follows:

$$(1 + n^{2}\cos \chi) = 1 + \frac {n}{v} + \frac{\omega'}{\omega}
\left(\frac{n}{v} + n^{2} \right) \eqno(14)$$
and

$$1 - nv\cos\chi = -\frac {\omega'}{\omega} (1 + nv).\eqno(15)$$

Formulas (14) and (15) will be used later in determination of the cross section.
From eq. (13) folows the relation between frequences $\omega$ and $\omega'$, if the initial
frequence $\omega'$ is emitted in the direction $\alpha = \pi$. Using

$$ \frac {1}{vn} \leq \cos\chi \leq  1,\eqno(16)$$
we get

$$\omega \geq \omega'\frac {vn+1}{vn-1}; \quad
vn \geq 1;  \quad \alpha = \pi\eqno(17)$$
and we see that when $nv \to 1 $, then, $\omega/\omega' \to \infty $.

\section{The existence of the \v CCE}

The \v CCE exists if and only if a charged
particle moves in a dielectric medium with the speed which is faster then the
speed of light in this medium. For velocities lower then the speed of light
in the medium, there is no \v CCE. Let us proof it.

In general, we have:

$$p - k = p' + k',\eqno(18)$$
or, after squaring the last equation, we have with $p^2 = m^2$ and
$k^2 = 0$ in the laboratory system where dielectric medium is in rest:

$$-(E\omega - {\bf p}\cdot {\bf k}) = E'\omega' -
{\bf p}'\cdot {\bf k}'.\eqno(19)$$

Then, with  ${\bf p } = E{\bf v}$ and $|k| = n\omega$ and $v' \approx v$ we have:

$$ - \frac {E\omega}{E'\omega'} = \frac{1 - vn\cos\alpha}{1 -
vn\cos\chi} < 0\eqno(20)$$

Now, we can decide two cases:

$$I: (1 - vn\cos\alpha) < 0; (1 - vn\cos\chi)  > 0,  \eqno(21)$$

$$II:    1 - vn\cos\alpha) > 0; (1 - vn\cos\chi)  < 0.\eqno(22)$$

If we now apply the \v Cerenkov conditio $vn >1$, we see that the first and the second  possibilities  can be true
for some angle $\alpha$ and $\chi$. On the other hand,  if we suppose velocities which are smaler
than the velocity of ligh in a medium,or, $vn < 1$, then, neither the first condition I, nor the second condition II can be
true for  $\alpha$ and $\chi$. So,  only for velocities which are greater than velocity of light in medium the two-photon
process can occur. So the denotation \v Cerenkov-Compton effect is appropriate.

It means, in other words,  that the conditions for the existence of the \v CCE
is the same as the condition of the existence
of the pure \v Cerenkov effect.

\section{The cross-section of the \v Cerenkov-Compton effect}

We shall use the notion cross-section although there is no targed. However, this notion has also a meaning of the
probability of generation of two photons of the specific parameters and from this point of view the cross-section is physically meaningful.

The cross-section of the \v CCE can be easily evaluated
in the so called Mandelstam variables $s, t, u$ which are defined,
as it is well
known, by the following relations in vacuum [14]:

$$s = (p + k)^{2} = m^{2} + 2p'k'\eqno(23)$$

$$t = (p - p') = -2kk'\eqno(24)$$

$$u = (p - k') = m^{2} -2p'k,\eqno(25)$$
where the right side of equations concerns the Compton situation
$p+k = p' + k'$. In the case
of the \v Cerenkov-Compton effect when $p = k + k' + p'$, we see that
the this relation can be obtained from the former relation,
by transformation
$k \to -k$. In such a way the Mandelstam parameters follows from
eqs. (23)--(25) in the form:

$$s = (p - k)^{2} = m^{2} + 2p'k'\eqno(26)$$

$$t = (p - p') = 2kk'\eqno(27)$$

$$u = (p - k') = m^{2} + 2p'k .\eqno(28)$$

The equations (26)--(28) are valid in vacuum and are invariant in all inertial systems.
However if we
consider the process in medium then, the invariance is not valid because
photon four-vector is transformed according to the specific transformation
appropriate to the medium. So, we will use the laboratory system where
dielectric medium is at rest and where $k^{2} = 0$. The derived formulas are valid for the laboratory system.

We express the last equations in terms  of the four-momenta
of electron and photon $(E, {\bf p}), (\omega, {\bf k})$.
So, in the laboratory system we have:

$$s =  m^{2} + 2E'\omega'(1 - v'n\cos \alpha)\eqno(29)$$

$$t =  2\omega\omega'(1 - n^{2}\cos\varphi)\eqno(30)$$

$$u =  m^{2}  + 2E'\omega(1 - nv'\cos\chi) .\eqno(31)$$

The cross-section has the following general form in the Mandelstam variables. [14].

$$d\sigma = \frac{8\pi r_{e}^{2}m^{2}dt}{(s-m^{2})^{2}}
\left\{\left(\frac {m^{2}}{s-m^{2}} + \frac {m^{2}}{u-m^{2}}\right)^{2} +
\right.$$

$$\left.
\left(\frac {m^{2}}{s-m^{2}} + \frac {m^{2}}{u-m^{2}}\right) -
\frac {1}{4}\left(\frac {s-m^{2}}{u-m^{2}} + \frac {u-m^{2}}{s-m^{2}}
\right)\right\} .\eqno(32)$$

Now, let us determine the $d\sigma$ of \v CCE for special situation where $\alpha = \pi$.

In this case $(E \approx E')$

$$s \approx  m^{2} +2E\omega'(1 + v n)\eqno(33)$$

$$t =  2\omega\omega'(1 + n^{2}\cos\chi)\eqno(34)$$

$$u \approx  m^{2}  + 2E\omega(1 - nv\cos\chi) ,\eqno(35)$$
or,

$$s - m^{2} \approx 2E\omega' (1 + nv) .\eqno(36)$$

$$t =  2\omega\omega'\left(1 + \frac{n}{v}\right) + 2\omega'^{2}
\left(\frac{n}{v} + n^{2}\right); \quad
dt|_{\omega' = const} = 2\omega'\left(1 + \frac {n}{v}\right) d\omega .
\eqno(37)$$

$$u -  m^{2} \approx - 2E\omega'(1 + nv) .\eqno(38)$$

For differential cross-section we then have with $r_{e} = e^{2}/m$:

$$d\sigma = \frac {2\pi e^{4}}
{E^{2}\omega'}
\frac {1 + n/v}{(1 + nv)^{2}} d\omega .\eqno(39)$$

Since $e^{2}$ has dimension energy $\times$ length, then the dimension
of $d\sigma$ is ${\rm length}^{2}$.
We use term cross-section, although there is no target. It means that
the physical meaning of this term is to be considered in terms of the probability of the process.

In case, we do not use the specification $\alpha = \pi$,
we get very complicated
equations for $t$ and $dt$. Also the final cross section formula is very complicated
and it means it is not suitable for experimental verification. In experiment,
simple formulas are more attractive, than complicated ones.

The experimental content of the last formula is as follows. To given
$\omega'$ which is emitted in angle $\alpha = \pi$, and detected by the
pixel detector placed in the direction of $\alpha = \pi$ there exist photon
with frequency $\omega$ which can be detected by the detector placed in the
direction $\chi$ given by equation (13). So, Using the pixel detectors,
the \v CCE is possible detect with the cross-section given by the last
formula. Let us remark that emission of photons in the direction
$\cos\chi > 1/nv$ was observed in CERN as a byproduct of the ordinary \v Cerenkov effect.

Let us still remark that the situation with photon moving at the opposite direction to the electron motion
has crucial heuristical meaning which has no analogy in particle physics. In particle physics
only collison with targed can produce  backscattering. Here we see that without scattering centers
the photons can be produced at the opposite direction.

\section{Discussion}

We have seen in the preceding chapters that the so called \v Cerenkov-Compton
effect can be rigorously formulated and solved in QED. We have used the
analogy with the Compton effect because the ordinary \v Cerenkov effect
can be determined from the Feynman diagram with the emission of one photon [4],
and the Compton diagram is only generalization to the situation with
two photons. At the same time it is necessary to say it was not possible
to use the theory of the double Compton effect [15]
because the initial state of
of the double Compton effect is $\gamma + e$, while the initial state of
the \v CCE is single electron. The \v CCE substantially differs from the ordinary \v Cerenkov effect
because it cannot be explained by any mechanical model being an effect of quantum electrodynamics.

We have proved that in  QED the \v Cerenkov-Compton effect occurs
as the integral part of every ordinary \v Cerenkov process.
While the \v Cerenkov effect is described in textbooks of
electrodynamics and monographies, articles and preprints, and it is observed
in all particle laboratories, the \v Cerenkov-Compton
effect is still not considered as an effect of experimental relevance.

However, this effect was yet probably observed in particle laboratories and no
appropriate attention was devoted to it. The situation is an analogue of
the situation when Pierre and Marie Curie observed \v Cerenkov
radiation in vessel with the radiaoactive salts and they were not aware
that it was the  \v Cerenkov radiation. Also Mallet [16]
in 1926-1929 during experiments with the gamma rays interacting with water observed the \v Cerenkov
radiation but he was not able to define this effect as the new phenomenon.

In CERN, the particle identification system is proposed for the
LHC-B experiment. It consists of RICH detectors with three materials. The
photodetectors under development are multipixel hybrid photodiodes, which
allow high single photon detection and high spatial resolution. According
to CERN report [17], there are background of photons which is in this report
interpreted as Rayleigh scattering of the \v Cerenkovian photons.
It is well known that the Rayleigh
scattering is an elastic scattering of photons on nonhomogenities and
density fluctuations of a medium leaving the electron energy unchanged.The cross section of this process
varies as the inverse fourth power of the photon wavelength, or, [18]

$$\sigma \propto \frac {1}{\lambda^4}.\eqno(40)$$

The transition radiation is not considerred because it can be neglected.
We of course suppose that also the process of \v Cerenkov-Compton
photons contributes to the background and at present time the two
kinds of photons are not distinguished. The colour photographies
of the \v Cerenkov radiation informs us  that the \v Cerenkov circles are
blue and the background inside of the circle(s) is not black [19].

We know that the pure \v Cerenkov  radiation is linearly polarized in the
plane defined by the direction of observation and path of particle motion.
The polarization of the Rayleigh photons probably differs
from the polarization of the \v Cerenkov, or, the \v Cerenkov-Compton photons.
So, if the background is formed by the Rayleigh photons and
the \v Cerenkov-Compton photons, it can be in principle possible
spectroscopically to distinguish both groups of
photons and to verify the existence of the \v CCE.
The second possibility how to register the \v Cerenkov-Compton photons is to
use absolutely homogenous medium at very low temperature. Then,
the Rayleigh scattering does not occur and we can observe only the \v CCE.

However, if in experiment it will be proved that background photons
are not from \v CCE, then it can be considerred as a crucial puzzle, because
the \v CCE is allowed by QED. This puzzle will open evidently new view on the \v CCE.

We hope that sooner or later the \v Cerenkov-Compton effect will
be again verified and investigated in the laboratories  of particle physics,
where the \v Cerenkov effect plays the substantial role in the experiment.


\begin{thebibliography}{99}
\bibitem{1}  Frank I. M. 1968 {\it Journal of Nuclear Physics} {\bf 7}  No. 5
 1100. (in Russian).
\bibitem{2}  Frank I. M. and Tsytovich V. N. 1980
{\it Journal of Nuclear Physics} {\bf 31} No. 4 974. (in Russian).
\bibitem{3}  Jelley J. V. 1958 {\it \v Cerenkov  Radiation},
(Pergamon Press, London).
\bibitem{4} Di Giacomo A., Paffuti G. and Rossi P.
1994 {\it Selected problems in theoretical physics} (World Scientific Publishing Co.
Pte. Ltd. Singapore) Problem 93 p. 304.
\bibitem{5} \v{C}erenkov P. A., C. R. 1936 {\it Acad. Sci.}
(USSR) {\bf 3}  413.
\bibitem{6} Tamm I. E. and  Frank I. M. 1937
{\it Dokl. Akad. Nauk USSR} {\bf 14} 107.
\bibitem{7} Schwinger J.,  Tsai W. Y. and  Erber T. 1976
{\it Ann. Phys.} (NY) {\bf 96}  303.
\bibitem{8}  Pardy M. 1983 {\it Phys. Lett.} {\bf 94} A  30.
\bibitem{9} Pardy M. 1983 {\it J. Phys. G: Nucl. Phys.} {\bf 9} 853.
\bibitem{10} Pardy M. 1989 {\it Phys. Lett.} A~{\bf 134} No. 6 357.
\bibitem{11} Pardy M. 1995 {\it Int. J. Theor. Phys.} {\bf 34} No. 6  951.
\bibitem{12} Pardy M. 1994 {\it Phys. Lett.} B {\bf 325} 517.
\bibitem{13} Pardy M. 1994 {\it Phys. Lett.} B {\bf 336}  362.
\bibitem{14} Berestetskii V. B., Lifshitz E. M. and  Pitaevskii L. M. 1989
{\it Quantum electrodynamics} 3rd ed. Landau and Lifshitz Course of Theoretical
Physics  Vol. 4 (Moscow, Nauka) (in Russian).
\bibitem{15} Mandl F., Skyrme T. H. R. 1952 {\it Proc. Roy. Soc.}  A {\bf 215}
 497.
\bibitem{16} Mallet M. L., C. R. 1926 {\it Acad. Sci.} (Paris) {\bf 183}  274. ;
ibid.  1928 {\bf 187} 222.; ibid.  1929 {\bf 188} 445.
\bibitem{17} Forty R. 1996 {\it CERN-PPE/96-176}, {\bf 4} September.
\bibitem{18} Jackson J. D. 1998 {\it Classical Electrodynamics}, 3rd ed.
(John Wiley
and Sons, Inc. New York).
\bibitem{19} {\it CERN COURIER} 1998 {\bf 38} No. 9 December p. 7  and
title page.
\end{thebibliography}
\end{document}